\documentclass[twocolumn,showpacs,preprintnumbers,amsmath,amssymb]{revtex4}
\usepackage{amsmath}
\usepackage{graphicx}
\usepackage{dcolumn}
\usepackage{bm}

\begin{document}
\preprint{APS/ Manuscript No. AL10337/Ou }
\title{Coherence Induced by Incoherent Pumping Field and Decay Process in Three-level $\Lambda$ Type Atomic System}
\author{Bao-Quan  Ou}
  \email{bqou@nudt.edu.cn}
\author{Lin-Mei Liang}
 \email{nmliang@nudt.edu.cn}
\author{Cheng-Zu Li}
\affiliation{Department of Physics, Science College, National
University of Defense Technology,  Changsha, 410073, People's
Republic of China .}
\date{\today}
\begin{abstract}
Following the method of Victor V. Kozlov \emph{et al.}[PhysRevA. 74.
063829],we inspect the atomic coherence induced by incoherent pump
and spontaneous decay process in $\Lambda$ type three-level atomic
system with degenerated lower duplicate levels. The system shows a
coherent population trapping state and multi-steady states
characteristic in different conditions. Interestingly, we derived
two kinds of different steady states generated by the system in
different sets of pumping and decaying parameters, the "robust"
steady state and the "weak" steady state, which exhibit stable or
unstable characteristics under the action of pump field and vacuum
reservoir. These two kinds of steady states help to understand the
coherence excitation mechanism, and will promise fruitful
applications to atomic coherence and interference in quantum optics.
\end{abstract}
\maketitle

\section{\label{sec:Introd}Introduction}
Quantum coherence effects such as coherent population
trapping(CPT)\cite{CPT}, electromagnetically induced
transparency(EIT)\cite{EIT}, lasering without
inversion(LWI)\cite{LWI}\cite{LWI2} and etc, are extensively studied
in quantum optics\cite{QuantOpt}\cite{QuantOpt2}. However most of
the study on atomic coherent effect focus on coherent process as the
coherent pump procedure, the possibility of atomic coherence induced
by incoherenct process was not recognized until late in the middle
of last decade\cite{Zhu95}. Since then, a lot of new schemes on
application of coherent effect based on the incoherent character of
atomic system are proposed , for example, quenching spontaneous
emission\cite{QuenchSp}\cite{QuenchSp2}, lasing without
inversion\cite{PRA063829} and so on. In fact the coherent process as
the coherent pump is not a unique condition to produce the upper
quantum coherent effects. For instance, in the V type three level
system, coherence can be produced by incoherent process, i.e., the
decay process and incoherent pump\cite{PRA063829}.

Following the method proposed in the Victor V. Kozlov's
paper\cite{PRA063829}, we discuss a different situation, the
$\Lambda$ type three level system. Here the $\Lambda$ system also
shows coherent population trapping feature and the multi-steady
state character. And we find out an interesting phenomenon that in
the coherent population trapping state, levels population are only
related to the incoherent pumping rates, totally unrelated to the
decay rate, which is different from the situation in the $V$ type
system, where the pumping and decay rates are relevant. Thus we can
adjust the proper pumping rate parameter to change the level
population from almost zero population state to nearly full
population state or vice versa, which is similar to the coherent
pumping three-level $\Lambda$ system to CPT state\cite{QuantOpt}.
What's more interesting is that under one special set of pumping
parameters, we produce a "robust" steady state, which remains its
original form under the interaction of pumping field and decaying
process.

In this paper we also inspect the population transferring picture
and atomic coherence excitation mechanism in detail, and find that
coherence is produced by the destructive interference of incoherent
pump process and decay process. Under specific conditions, the decay
process can be totally suppressed by incoherent pump process, as a
result, the "robust" steady state is produced. On the other hand,
when the pumping field is switched off, coherence can also be
generated by spontaneous emission decay procedure, and the final
state of atomic system comes to the "weak" steady state, which shows
unstable property under nonvanishing pump driving condition. The
peculiar nature of the $\Lambda$ type three-level atomic system will
promise fruitful applications in quantum optics.

This paper is organized as follows: in Sec.\ref{sec:MasEq}, we
derive the motion equations for density-matrix of the three level
$\Lambda$ atomic system. In Sec.\ref{sec:Sol}, we derive solutions
to the atomic motion equations under two kind of conditions, the
general steady state situation in Sec.\ref{sec:robust} and the
specific initial conditions in Sec.\ref{sec:weak}, where the
interesting character of the atomic system will be revealed. Finally
in Sec.\ref{sec:Diss} we discuss the potential applications for the
system and set our conclusion.

\section{\label{sec:MasEq}Master Equation of Three Level $\Lambda$ Atomic system}
\begin{figure}
\scalebox{0.75}{\includegraphics[width=\columnwidth]{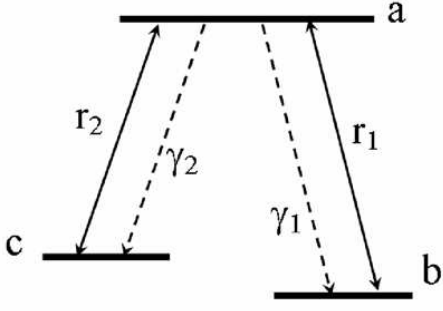}}
\caption{\label{fig:lambdasys} Scheme of a three-level $\Lambda$
type atomic system with closely doublet lower levels. $r_{1}$and
$r_{2}$ are incoherent pump rates; $\gamma_{1}$ and $\gamma_{2}$ are
decay rates.}
\end{figure}

We are discussing a $\Lambda$ type three level atomic system shown
in Fig.\ref{fig:lambdasys}, where the upper state $|a\rangle$ is
connected to two lower closely spaced states $|b\rangle$ and
$|c\rangle$ by dipole-allowed transition. The incoherent pump field
$R$ drives the two lower states to the same upper state and back.
The upper state decays through two different paths to lower states.

Firstly we will derive the motion equations of the density-matrix
elements for the atomic system. The interaction hamiltonian of the
system in the interaction picture is shown in Eq.(\ref{eq:Hamt}):
\begin{equation}
H = V_\gamma+V_R
 \label{eq:Hamt}.
\end{equation}
The first term on the right side of Eq.(\ref{eq:Hamt}), $V_\gamma$
is the interaction with the reservoir of vacuum oscillators. The
specified hamiltonian of this term reads as Eq.(\ref{eq:hvac}).
\begin{eqnarray}
V_\gamma =\hbar\sum_{k}[g_{k1}a_{k}^{\dag}|b\rangle\langle{a}|e^{-i\Delta_{k1}t}+g_{k1}^{*}a_{k}|a\rangle\langle{b}e^{i\Delta_{k1}t} \nonumber\\
 +g_{k2}a_{k}^{\dag}|b\rangle\langle{c}|e^{-i\Delta_{k2}t}+g_{k2}^{*}a_{k}|c\rangle\langle{b}e^{i\Delta_{k2}t}]
\label{eq:hvac},
\end{eqnarray}
where $g_{k1,k2}$ are the coupling constants between the \emph{k}th
vacuum mode and the atomic transitions from level $|a\rangle$ to
level $|b\rangle$ (from level $|a\rangle$ to level $|c\rangle$).
Without losing the generality, we take the coupling constants
$g_{k1,k2}$ the real numbers. The detunings
$\Delta_{k1}=\omega_{ab}-\nu_{k},\Delta_{k2}=\omega_{ac}-\nu_{k}$
are the difference between transitions frequencies $\omega_{ab}$,
$\omega_{ac}$ and the vacuum mode $\nu_{k}$.  And
$\hat{a}_k$$(\hat{a}^{\dag}_{k})$ is the annihilation(creation)
operator of a photon in the \emph{k}th vacuum mode, which obey the
bosonic commutation rule:
$[\hat{a}_{k},\hat{a}^{\dag}_{k'}]=\delta_{kk'}$.

Another incoherent process in Fig.\ref{fig:lambdasys} is the
incoherent pumping process, which is described in the second term
$V_R$ on the right side of Eq.(\ref{eq:Hamt}), the specified
hamiltonian of incoherent field $R$ takes the following form of
Eq.(\ref{eq:Pfield}):
\begin{equation}
V_R=\mu_{ba}R^{*}\sigma_{1}^{-}+\mu_{ab}R\sigma_{1}^{\dag}+\mu_{ca}R^{*}\sigma_{2}^{-}+\mu_{ac}R\sigma_{2}^{\dag}
 \label{eq:Pfield},
\end{equation}
where the dipole matrix elements $\mu_{ab}$ and $\mu_{ac}$ are the
coupling constants for the incoherent field $R$ coupling with the
atomic system. The following relationship shows the incoherent
feature of the pumping field $R$:
\begin{equation}
\langle R^{*}(t)R(t') \rangle=\mathcal {R}\delta(t-t')
 \label{eq:Pfdelta}.
\end{equation}

The Liouville equation for the atomic system, incoherent pumping
field and vacuum reservoir is:
\begin{equation}
\dot{\rho}_T=-\frac{i}{\hbar}[H,\rho]
 \label{eq:master},
\end{equation}
where $\rho_T$ is the density operator for total system of the atom,
pumping field and vacuum reservoir, and the hamiltonian \emph{H}
consists of the vacuum induced decaying part and the incoherent
pumping part. We first formally integrate the Eq.(\ref{eq:master}),
and then substitute it back to the Eq.(\ref{eq:master}), neglecting
terms higher than second order:
\begin{equation}
{\dot{\rho}}_{T}=-\frac{i}{\hbar}[H(t),\rho_{T}(0)]-\frac{1}{{\hbar}^2}\int^t_0
[H(t),[H(t'),\rho_{T}(t')]]dt'
 \label{eq:fint},
\end{equation}
where $t=0$ is chosen as the initial time. Note that the hamiltonian
$H$ contains two separated processes, the decaying one and the
pumping one, we can treat them independently in the upper
Eq.(\ref{eq:fint}), and then joint them together to get the motion
equation for the concerned atomic system. Take the decaying part one
for instance, we take the routine Weisskopf-Wigner approximation,
assuming that the reservoir is large enough and the couple to the
atomic system is very weak, then the density operator $\rho_T$ for
the total system can be factorized into direct product of the atomic
density operator $\rho$ and the density operator of reservoir
$\rho_r$ at all time: $\rho_T=\rho\otimes\rho_r$. With this
assumption, the motion equation for the atomic system is:
\begin{eqnarray}
\dot{\rho}=Tr_{R}[\dot{\rho}_T]=-\frac{i}{\hbar}Tr_{R}[H(t),\rho_{T}(0)] \nonumber\\
-\frac{1}{{\hbar}^2}Tr_{R}\int^t_0
[H(t),[H(t'),\rho_{T}(t')]]dt'
 \label{eq:AtMasterEq},
\end{eqnarray}
where $Tr_{R}$ denotes trace over the vacuum reservoir.

After long and complex calculation, we obtain the following set of
equations for the atomic density matrix
elements\cite{QuantOpt}\cite{PRA063829}:
\begin{eqnarray}
\label{eq:rhoaa}
\dot{\rho}_{aa}=-(\gamma_1+\gamma_2+r_1+r_2)\rho_{aa}+r_1\rho_{bb}+r_2\rho_{cc}\nonumber\\
+p\sqrt{r_1r_2}(\rho_{bc}+\rho_{cb})\\
 \dot{\rho}_{cc}=\gamma_2
\rho_{aa}+r_2(\rho_{aa}-\rho_{cc})-\frac{1}{2}p\sqrt{r_1
r_2}(\rho_{bc}+\rho_{cb})\\
\dot{\rho}_{bb}=\gamma_1
\rho_{aa}+r_1(\rho_{aa}-\rho_{bb})-\frac{1}{2}p\sqrt{r_1
r_2}(\rho_{bc}+\rho_{cb})\\
\label{eq:enclose}
\rho_{aa}+\rho_{bb}+\rho_{cc}=1\\
\label{eq:rhoab}
\dot{\rho}_{ab}=-\frac{1}{2}(\gamma_1+\gamma_2+2r_1+r_2)\rho_{ab}-\frac{1}{2}p\sqrt{r_1r_2}\rho_{ac}\\
\label{eq:rhoca}
\dot{\rho}_{ac}=-\frac{1}{2}(\gamma_1+\gamma_2+r_1+2r_2)\rho_{ac}-\frac{1}{2}p\sqrt{r_1r_2}\rho_{ab}\\
\dot{\rho}_{bc}=-\frac{1}{2}(r_1+r_2)\rho_{bc}+p\sqrt{\gamma_1\gamma_2}\rho_{aa}\nonumber\\
+\frac{1}{2}p\sqrt{r_1r_2}(2\rho_{aa}-\rho_{bb}-\rho_{cc})+i\Delta\rho_{bc}
\label{eq:rhocb}.
\end{eqnarray}

Eq.(\ref{eq:enclose}) expresses the conservation of probability for
the closed $\Lambda$ system. The decay and pump constants are
defined as follow: $\gamma_{1,2}\equiv 2 \pi
D(\omega)g^{2}_{\omega}$ and
$r_{1,2}\equiv2(\mu^{2}_{ab,ac}/\hbar^{2})\mathcal {R}$. Here
$g_{k1,k2}=g_{\omega}$ are calculated at the transition frequency
$\omega=\omega_{ac}$. The detuning is defined as:
$\Delta=\omega_{ac}-\omega_{bc} $, and in the following we set
$\Delta=0$. The density of states is:
$D(\nu_k)={V\nu^{2}_{k}}/{\pi^2 c^3}$, with $V$ the quantization
volume and $c$ the vacuum speed of light, and we take the
approximation: $D(\nu_k)=D(\omega)$ when calculate frequency
integral. The $p$ factor in the
Eq.(\ref{eq:rhoaa})-Eq.(\ref{eq:rhocb}) is the alignment of the
dipole matrix elements, defined as Eq.(\ref{eq:pfactor}):
\begin{equation}
p\equiv\frac{\mu_{ac}\mu_{bc}}{|\mu_{ac}||\mu_{bc}|}
 \label{eq:pfactor},
\end{equation}
and it takes value from the set of $[-1,1]$, 1 for parallel dipole
moments, -1 for antiparallel, and 0 for orthogonal. For simplicity
we set $p=1$ in the following discussion.

\section{\label{sec:Sol} Solutions to Master Equations}
Solutions to the master equations
Eq.(\ref{eq:rhoaa})-Eq.(\ref{eq:rhocb}) is complicated, but we are
interested in the steady solutions, i.e., the long time evolution of
the system, to see the coherence induced by the combined interaction
of incoherent pumping and decaying process. In solving the master
equation for the atomic system, we first take the routine method of
steady state solution, assuming that all the density matrix elements
for the atomic system do not evolve with time in the long run. We
find that in the steady state the atomic system turns out to be a
coherent population trapping (CPT) state, as the result of coherent
pumping field driving a $\Lambda$ atomic system, and the CPT state
we derived here shows some interesting characters. The details are
shown in the section \ref{sec:robust}. Secondly, we solve the motion
equations for the atomic system under several kinds of initial
conditions, the solutions are analytic, from which we learn the
details about population transferring and coherence producing, the
content will be shown in the section \ref{sec:weak}.

\subsection{\label{sec:robust}General Steady Solutions to Master Equations: Coherent Population Trapping state}
Now let's discuss the general steady solutions to the master
equations Eq.(\ref{eq:rhoaa})-Eq.(\ref{eq:rhocb}). Letting the left
sides of the Eq.(\ref{eq:rhoaa})-Eq.(\ref{eq:rhocb}) equal to zero,
and solving the resulted equations, we have the following general
steady solutions
Eq.(\ref{eq:steadrhoaa})-Eq.(\ref{eq:steadrhobc})(note that we have
set $\Delta=0$ and $p=1$, the degenerated lower atomic states
assumption and the parallel pumping dipole moments situation):

\begin{eqnarray}
\label{eq:steadrhoaa}
\rho_{aa}=0,\\
{\rho}_{cc}=\frac{r_1}{r_1+r_2}\\
{\rho}_{bb}=\frac{r_2}{r_1+r_2}\\
 \label{eq:steadrhobc}
{\rho}_{bc}=-\frac{\sqrt{r_1r_2}}{r_1+r_2}
\end{eqnarray}

The condition leading to the unique solutions
Eq.(\ref{eq:steadrhoaa})-Eq.(\ref{eq:steadrhobc}) is that the
parameters set satisfied the following equation
Eq.(\ref{eq:parameter}):
\begin{eqnarray}
\label{eq:parameter}
 r_{2}\gamma_{1}+r_{1}\gamma_{2}-2\sqrt{r_{1}r_{2}\gamma_{1}\gamma_{2}}\neq{0}\nonumber\\
r_{1}r_{2}\neq{0}
\end{eqnarray}

It's interesting that the steady state solutions
Eq.(\ref{eq:steadrhoaa})-Eq.(\ref{eq:steadrhobc}) shows that the
upper level state $|a\rangle$ has no population, and the two lower
states $|b\rangle$ and $|c\rangle$ share the total population whose
amount is in proportion to the incoherent pumping rate $r_2$ and
$r_1$, respectively, and they have no relationship to the decaying
rates $\gamma_1$ and $\gamma_2$. This is the typical coherent
population trapping (CPT) state, and coherent trapping occurs due to
the destructive interference of incoherent pumping field and the
vacuum reservoir, as will be explained in the following text. The
peculiar feature of the lower state population indicates that no
matter how strongly the incoherent pumping field drives the system,
the upper level remains empty, the pumping field can only change the
lower state population. In order to inspect the characters of the
solutions Eq.(\ref{eq:steadrhoaa})-Eq.(\ref{eq:steadrhobc}) further,
let's switch to the dressed state picture. We define the following
"dark" and "bright" state:
\begin{eqnarray}
\label{eq:dress}
|D\rangle=\frac{\sqrt{r_2}|b\rangle-\sqrt{r_1}|c\rangle}{\sqrt{r_1+r_2}}\\
|B\rangle=\frac{\sqrt{r_1}|b\rangle+\sqrt{r_2}|c\rangle}{\sqrt{r_1+r_2}}.
\end{eqnarray}

The $|D\rangle$ state is the dark state and represents the CPT
state, while the $|B\rangle$ state is the bright state and it decays
rapidly. The dressed state density matrix elements and the
corresponding bare state density matrix elements have the following
transformation:
\begin{eqnarray}
\label{eq:denmatdress}
\rho_{DD}=\frac{{r_2}\rho_{bb}+{r_1}\rho_{cc}-2\sqrt{r_1r_2}\rho_{bc}}{{r_1+r_2}}\\
\rho_{BB}=\frac{{r_1}\rho_{bb}+{r_2}\rho_{cc}+2\sqrt{r_1r_2}\rho_{bc}}{{r_1+r_2}}\\
\rho_{DB}=\frac{{\sqrt{r_1r_2}}\rho_{bb}-\sqrt{r_1r_2}+(r_2-r_1)\rho_{bc}}{{r_1+r_2}}.
\end{eqnarray}

Then the evolution equations for the dressed state density matrix
elements read as:
\begin{eqnarray}
\label{eq:dressEvol}
{\dot{\rho}_{DD}}=\frac{(\sqrt{r_2\gamma_1}-\sqrt{r_1\gamma_2})^2\rho_{aa}}{{r_1+r_2}}\\
{\dot{\rho}_{BB}}=\frac{[(\sqrt{r_1\gamma_1}+\sqrt{r_2\gamma_2})^2+(r_1+r_2)^2]\rho_{aa}}{{r_1+r_2}}\nonumber\\
-(r_1+r_2)\rho_{BB}\\
{\dot{\rho}_{aa}} =-(r_1+\gamma_1+r_2+\gamma_2)\rho_{aa}+(r_1+r_2)\rho_{BB}\\
\label{eq:dressDB}
 {\dot{\rho}_{DB}}
=\frac{[(\gamma_1-\gamma_2)\sqrt{r_1r_2}-(r_1-r_2)\sqrt{\gamma_1\gamma_2})]}{{r_1+r_2}}\rho_{aa}\nonumber\\
-\frac{1}{2}(r_1+r_2)\rho_{DB}.
\end{eqnarray}

From Eq.(\ref{eq:dressEvol}) to Eq.(\ref{eq:dressDB}) it's clear
that the dark does not decay as time goes by, while the bright state
decay rapidly with rate $(r_1+r_2)$, the upper level population
$\rho_{aa}$ decay even fast with rate $(r_1+\gamma_1+r_2+\gamma_2)$,
and the dark-bright state coherence $\rho_{DB}$ decay with rate
$\frac{1}{2}(r_1+r_2)$, then it's evident that in the long run the
bright state population, the upper level population and the
coherence decay to zero, left the dark state population in a
nontrivial value, that is, the system will eventually stay in the
dark state.

Thus the steady state solution for the $\Lambda$ system driven by
incoherent pumping field and vacuum reservoir is a pure CPT state,
and it has many attractive characters, for example, the
"adiabatically" population transferring. If we start with the atom
in one of the lower states, the $|c\rangle$ state, for example, and
keep $r_1=0$, then the other lower state $|b\rangle$ will gradually
increase its population with pump rate $r_2$ finite, and eventually
the atom arrives at the $|b\rangle$ state. Considering the
parameters conditions Eq.(\ref{eq:parameter}), the pump rate $r_1$
and $r_2$ can not take zero value, we can set $r_1$ to be a very
small quantity, and the $|b\rangle$ state will ends up with almost
full population. In the dressed state picture, it's clear that the
atomic system remains in the "dark" state through out the process.
This procedure is like the "adiabatically" population transferring
process for CPT state in coherent driving the $\Lambda$ type
three-level atomic system\cite{QuantOpt}, but here we achieve the
same outcome by incoherent pumping field, and simply by changing the
pump rates, which is much faster and easier than that in the
coherent pumping driving situation.

In a special situation that the two pumping rates are equal:
$r_1=r_2=r$, the two lower states have equal population:
$\rho_{cc}=\rho_{bb}=\frac{1}{2}$, and the coherence between them
reaches the maximal value:${\rho}_{cb}=-\frac{1}{2}$. This special
steady state has a very interesting character, for instance, it does
not change as time goes on, further more, for arbitrary pumping and
decaying parameter (the two equal pump rate $r$ takes arbitrary
nonvanishing value), it will always stays at the maximal coherent
state, which shows a very stable character, we would call it the
"robust" steady state. The matrix of the "robust" steady state is:
\begin{eqnarray}
\label{eq:matrix} M=\left[
\begin{array}{cc}
 \rho_{bb} &\rho_{bc} \\
 \rho_{cb} &\rho_{cc}
   \end{array}
   \right]=\left[
   \begin{array}{cc}
 \frac{1}{2} &-\frac{1}{2}\\
-\frac{1}{2} & \frac{1}{2}
   \end{array}
   \right].
\end{eqnarray}
The matrix for shows that the stable feature for the state comes
from the coherent term: the off-diagonal elements of the matrix come
to their maximum value. If we express atomic level states in the
\emph{dual-rail} representation\cite{DualRail}, denote that level
state $|b\rangle$ has population as $|01\rangle$ state, and level
state $|c\rangle$ has population as $|10\rangle$ state, then the
"robust" steady state is:
\begin{equation}
\label{eq:RobustState}
|\psi\rangle=\frac{1}{\sqrt{2}}(|01\rangle-|10\rangle).
\end{equation}
It is clear that the "robust" steady state is a superposition of two
level states, and the amplitude of the two level states interference
comes to maximum value $-\frac{1}{2}$, corresponding to the maximum
coherent superposition state. In the following text we will see that
the coherence results from the interference of incoherent pumping
field and the spontaneous decay process, and due to different
contribution provided by the two processes, some more steady states
other than the "robust" steady state will appear.

\subsection{\label{sec:weak}Steady State Solutions to Master Equations Under Special Conditions }
In the upper section \ref{sec:robust} we have derived the general
steady solution to the master equations for the $\Lambda$ atomic
system, and had a CPT state and a "robust" steady state. However the
upper derivation is under the restriction of Eq.(\ref{eq:parameter})
for pumping and decaying parameters. If the parameters do not
satisfy Eq.(\ref{eq:parameter}), the upper solution is not unique
any more, and the master equations will have infinite solutions. In
the following, we will discuss some interesting situations leading
to many particular solutions to the master equations
(\ref{eq:rhoaa})-(\ref{eq:rhocb}).

As before, we set $p=1$ and $\Delta=0$, but what's more important,
we set pump rates and decay rates pairwise equal: $r_1=r_2=r$ and
$\gamma_1=\gamma_2=\gamma$, which does not satisfy the parameters
relationship Eq.(\ref{eq:parameter}). As a result, solutions to the
master equations (\ref{eq:rhoaa})-(\ref{eq:rhocb}) are different
from equations (\ref{eq:steadrhoaa})-(\ref{eq:steadrhobc}), and we
should solve the master equations according to different initial
conditions. The equations (\ref{eq:rhoaa})-(\ref{eq:rhocb}) become
considerably simplified for the conditions $r_1=r_2=r$ and
$\gamma_1=\gamma_2=\gamma$, and it is easy to find the following
"nondecaying combination"\cite{PRA063829} for the density matrix
elements:
\begin{equation}
\frac{d}{dt}(\rho_{aa}+\rho_{bc}+\rho_{cb})=0
 \label{eq:nondecay}.
\end{equation}.

The Eq.(\ref{eq:nondecay}) suggests that
$(\rho_{aa}+\rho_{bc}+\rho_{cb})$ does not evolve and then we can
derive the combination:
\begin{equation}
\rho_{aa}+\rho_{bc}+\rho_{cb}=C_0
 \label{eq:nondecay2}.
\end{equation}
where $C_0$ is the integration constant, it locates at the the
 $[-1,1]$ segment and it is determined by initial conditions
of the system, different values of $C_0$ corresponds to different
initial conditions. We will discuss several kinds of value of $C_0$
with corresponding initial conditions, as shown in table.
\ref{tab:table1}.
\begin{table}
\caption{\label{tab:table1}Different initial conditions correspond
to different values of $C_0$. }
\begin{tabular}{|c|c|c|}
\hline
Case & Initial condition & Value for $C_0$\\
\hline
& $\rho_{aa}=0,\rho_{bb}=1,\rho_{cc}=0,\rho_{bc}=\rho_{cb}=0 $ & \\
\cline{2 - 2}
$I$ & $\rho_{aa}=0,\rho_{bb}=0,\rho_{cc}=1,\rho_{bc}=\rho_{cb}=0$ & 0\\
\cline{2 - 2}
& $\rho_{aa}=0,\rho_{bb}=\frac{1}{2},\rho_{cc}=\frac{1}{2},\rho_{bc}=\rho_{cb}=0 $& \\
\hline
& $\rho_{aa}=0,\rho_{bb}=\frac{1}{2},\rho_{cc}=\frac{1}{2},\rho_{bc}=\rho_{cb}=\frac{1}{2} $ & \\
\cline{2 - 2}
\raisebox{1.5ex}[0pt]{$II$}& $\rho_{aa}=1,\rho_{bb}=0,\rho_{cc}=0,\rho_{bc}=\rho_{cb}=0 $& \raisebox{1.5ex}[0pt]{1} \\
\hline $III$
&$\rho_{aa}=0,\rho_{bb}=\frac{1}{2},\rho_{cc}=\frac{1}{2},\rho_{bc}=\rho_{cb}=-\frac{1}{2}
$& -1 \\
\hline
\end{tabular}
\end{table}

The first type of initial condition comes to $C_0=0$, which contains
three kinds of different level population and coherence, we denote
it the case $I$, while the second type of initial condition comes to
$C_0=1$, there are two kinds of level population and coherence, and
we denote it the case $II$. The third kind of initial conditions are
listed in case $III$, which corresponds to $C_0=-1$, and it is the
so-called maximal coherent situation.

\begin{figure*}
\textbf{(a)}
\scalebox{0.9}{\includegraphics[width=\columnwidth]{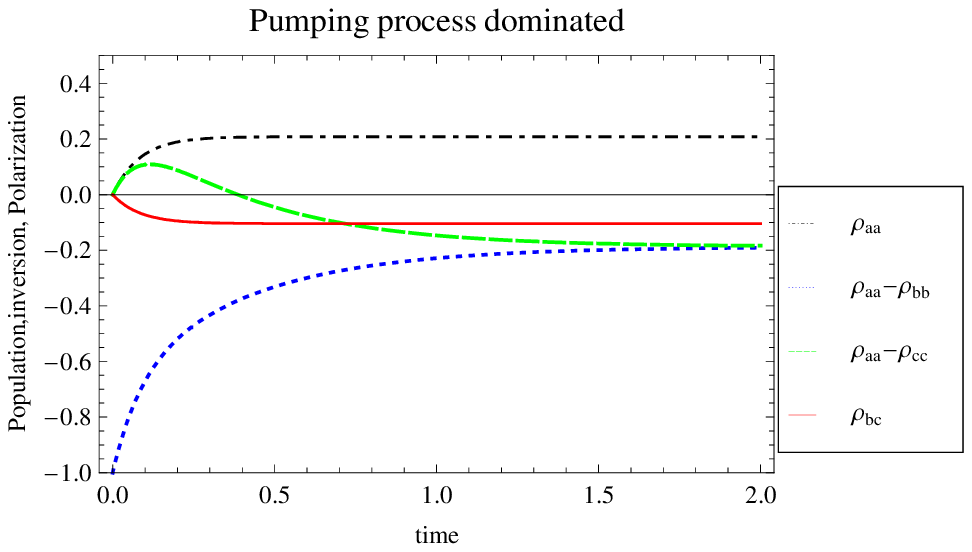}}
\textbf{(b)}
\scalebox{0.9}{\includegraphics[width=\columnwidth]{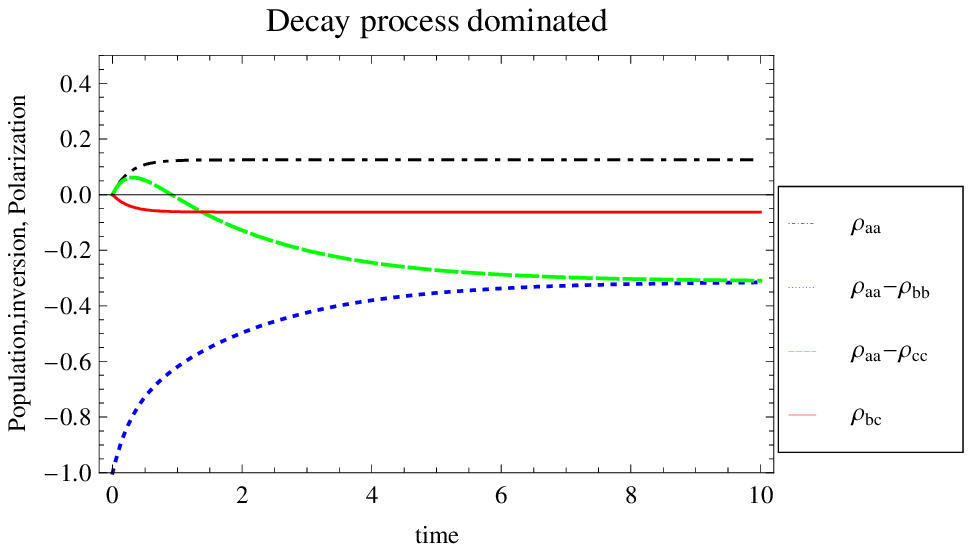}}
\caption{\label{fig:eqsolve1}Under initial condition case $I$
:$\rho_{aa}=0,\rho_{bb}=1,\rho_{cc}=0,\rho_{bc}=\rho_{cb}=0 $ ,
upper level population $\rho_{aa}$, population inversion
$\rho_{aa}-\rho_{bb}$,$\rho_{aa}-\rho_{cc}$, coherence
$\rho_{bc}$.Parameters are $r_1=r_2=r,\gamma_1=\gamma_2=\gamma$,Time
is measured in units of $\gamma$.(a):no population inversion in the
steady state, pumping interaction dominated: $r=2.5\gamma$;(b): no
population inversion in the steady state, decaying mechanics
dominated: $r=0.5\gamma$}
\end{figure*}

With the help of Eq.(\ref{eq:nondecay2}), we simplify the
Eq.(\ref{eq:rhoaa})- Eq.(\ref{eq:rhocb}):
\begin{eqnarray}
\label{eq:simplerhoaa}
\dot{\rho}_{aa}=r(C_0+1)-(2\gamma+4r)\rho_{aa}\\
 \dot{\rho}_{cc}=(\gamma+\frac{3}{2}r-\frac{1}{2}r C_0)-(\gamma+\frac{5}{2}r)\rho_{cc}-(\gamma+\frac{3}{2}r)\rho_{bb}\\
\dot{\rho}_{bb}=(\gamma+\frac{3}{2}r-\frac{1}{2}r C_0)-(\gamma+\frac{5}{2}r)\rho_{bb}-(\gamma+\frac{3}{2}r)\rho_{cc}\\
\dot{\rho}_{bc}=(\gamma+\frac{3}{2}r)
C_0-\frac{1}{2}r-(\gamma+\frac{5}{2}r)\rho_{bc}-(\gamma+\frac{3}{2}r)\rho_{cb}\\
 \label{eq:simplerhocb}
\dot{\rho}_{cb}=(\gamma+\frac{3}{2}r)
C_0-\frac{1}{2}r-(\gamma+\frac{5}{2}r)\rho_{cb}-(\gamma+\frac{3}{2}r)\rho_{bc}\\\nonumber
\end{eqnarray}
It's much easier to solve these equations. Take an example to solve
equations (\ref{eq:simplerhoaa})-(\ref{eq:simplerhocb}) in case $I$:
$\rho_{aa}=0,\rho_{bb}=1,\rho_{cc}=0,\rho_{bc}=\rho_{cb}=0 $, that
is, initially there are no population in the upper level state
$|a\rangle$ and lower level state $|c\rangle$, all population are in
the lower level state $|b\rangle$, the analytical results are shown
in Fig.\ref{fig:eqsolve1}, where the upper level population
$\rho_{aa}$, the population inversion $\rho_{aa}-\rho_{bb}$,
$\rho_{aa}-\rho_{cc}$ and coherence $\rho_{bc}$ are plotted.

From Fig.\ref{fig:eqsolve1} it's clear that all levels population do
not vary with time in the long run, indicate that the atomic system
goes into steady state. For different sets of pumping and decaying
parameters, the atomic system goes into different steady states, as
show in figure (a)(where pumping rate is set to be greater than the
decay rate) and figure (b)(where decay rate is larger than pumping
rate). By inspecting all the solutions corresponding to the
situations listed in the table.\ref{tab:table1}, we find that the
atomic system steps into corresponding steady states for different
initial conditions and pumping and decaying parameters sets, thus
the system shows a multi-steady state character.

From Fig.\ref{fig:eqsolve1} we find that, the population inversion
function $\rho_{aa}-\rho_{bb}$ and $\rho_{aa}-\rho_{cc}$ are both
take values below zero, which indicates that there is no population
inversion situations appear. And this is the common character of
case $I$ after inspecting the other two initial conditions. But for
initial conditions in case $II$, population inversion will appear in
the steady states. It's worthy to point out that in one situation of
case $II$, initially the two lower level states equally share the
total population, and the coherence between them take the "positive
maximum" value $\frac{1}{2}$, the system will evolve into another
steady state for nonvanishing pump rate.

\begin{figure}
\begin{center}
\includegraphics[width=\columnwidth]{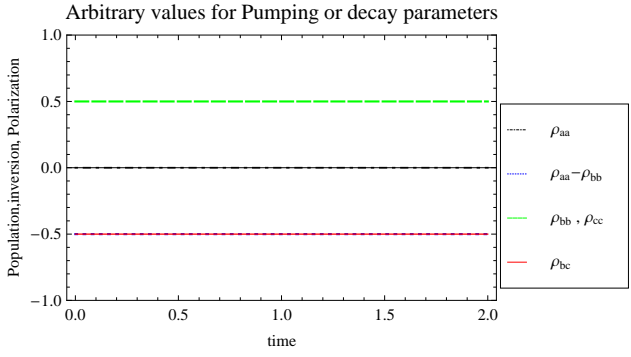}
\end{center}
\caption{\label{fig:maxcoh}Under initial condition case
$III$:$\rho_{aa}=0,\rho_{bb}=\frac{1}{2},\rho_{cc}=\frac{1}{2},\rho_{bc}=\rho_{cb}=-\frac{1}{2}
$ , upper level population $\rho_{aa}$, population inversion
$\rho_{aa}-\rho_{bb}$ and $\rho_{aa}-\rho_{cc}$,lower level
population $\rho_{bb}$ and $\rho_{cc}$, coherence $\rho_{bc}$. The
coherence function line is covered by population inversion function
lines. Parameters are $r_1=r_2=r,\gamma_1=\gamma_2=\gamma$,Time is
measured in units of $\gamma$. This is the maximal coherent state
situation, where level population does not change as time going by,
coherence does not change either.}.
\end{figure}

The most interesting solutions to the equations
(\ref{eq:simplerhoaa})-(\ref{eq:simplerhocb}) are achieved in case
$III$, where initially the two lower states have equal one half of
total population and the coherence between them was prepared to the
maximal value $-\frac{1}{2}$, it is the "maximal coherence"
generated in the system\cite{PRA063829}. From the results shown in
Fig.\ref{fig:maxcoh} we find out that this is the "robust" steady
state solution in section\ref{sec:robust}! It's clear that for
arbitrary sets of pumping $r$ and decay parameters $\gamma$, all
three level population $\rho_{aa}$,$-\rho_{bb}$ and $\rho_{cc}$ and
the coherence term $\rho_{bc}$ are staying in their initial values,
which shows a very stable characteristic.

Different from the general steady state solution, these solutions
corresponding to situations listed in table.\ref{tab:table1} are the
restrict analytic solutions with variable $t$, which will provide
much more system evolution details. Thus by inspecting the way of
level population transition and coherence acting in the $\Lambda$
system, we can reveal the mechanism of atomic coherence excitation
and preparation the "robust" steady state, and look for other
interesting state as the "weak" steady state.

From the results of the upper three kinds of situations represented
in Fig.\ref{fig:eqsolve1} - Fig.\ref{fig:maxcoh}, we can draw a
picture of level population transition. Take the result of
Fig.\ref{fig:eqsolve1} for instance, population starts from one of
the lower state, the $|b\rangle$ state, while the other two states
$|a\rangle$ state and $|c\rangle$ state are empty, and initially the
coherence $\rho_{bc}$ is zero, under the interaction of incoherent
pump field and vacuum decay process, population transfer to state
$|a\rangle$ and state $|c\rangle$. As shown in
Fig.\ref{fig:eqsolve1} (a), $|a\rangle$ state population grows up
from zero, after a time about $0.5\gamma$, it equals to that of
$|c\rangle$ state, and then it does not change with time, indicating
that the $|a\rangle$ state has stepped into steady state. The
population of the two lower levels are still changing, till time
equal to about $1.8\gamma$, state $|c\rangle$ and state $|b\rangle$
are arriving to steady state and their population are equal. As a
comparison, in Fig.\ref{fig:eqsolve1} (b), where the pump rate is
relatively weak: $r=0.5\gamma$, it costs much long time to reach
steady state for states $|b\rangle$ and $|c\rangle$, and this is the
same situation for other solutions in different initial conditions.

During the population transition period, the coherent term
$\rho_{bc}$ varies from initial value to steady value, for different
initial conditions, the coherence term $\rho_{bc}$ takes different
values in steady state, from positive values to negative ones, and
the varying laws are also different from each other. We would like
to inspect the varying law for coherence in a sophisticated way to
find out the mechanism of atomic coherence excitation. Depending on
different initial conditions shown in table.\ref{tab:table1}, the
forms of analytic solutions to coherence $\rho_{bc}$ are different
from each other, but for long time limit, they come to an identical
one as Eq.(\ref{eq:LTCoh}):
\begin{equation}
\label{eq:LTCoh}
 \rho_{bc}=\frac{(3r+2\gamma)C_0-r}{4(2r+\gamma)}
\end{equation}
where $C_0$ is the integration constant as defined in
Eq.(\ref{eq:nondecay2}), $r$ is the pump rate, $\gamma$ the decay
rate. Firstly we fix the value of decay rate $\gamma$ and discuss
the law of coherence term varies with pump rate. For initial
conditions of $C_0=0$, the coherence $\rho_{bc}$ takes negative
value, the absolute value of it become larger as pump rate $r$
increases; for initial conditions $C_0=1$, the coherence term
$\rho_{bc}$ takes positive value, but it will drops with pump rate
$r$ increases; for $C_0=-1$, coherence $\rho_{bc}=-\frac{1}{2}$,
remains unchanged. On the other hand, if we fixed the pump rate $r$
and inspect the varying law for coherence with decay rate, we will
get the contrary results. Combining different behaviors of the
coherence term $\rho_{bc}$ in Fig.\ref{fig:eqsolve1} -
Fig.\ref{fig:maxcoh}, we know that the coherence is induced by the
interference of incoherent pumping process and decay process, the
two processes contribute different sign to the coherence and
interfere destructively. Thus the coherence produces various of
steady state results according to different conditions, including
the general steady state solution Eq.(\ref{eq:steadrhobc}), in which
the contribution of decaying process was totally suppressed by
incoherent pumping process.

\begin{figure}
\begin{center}
\includegraphics[width=\columnwidth]{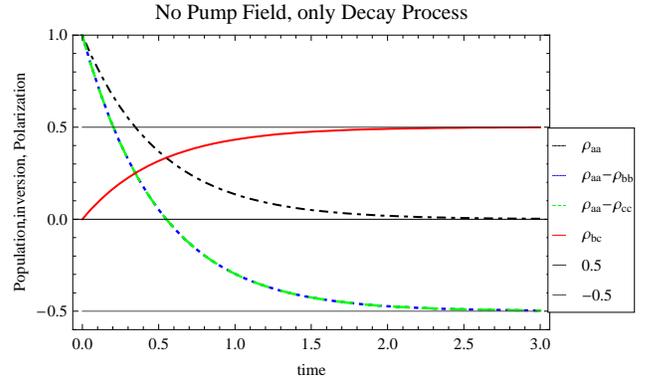}
\end{center}
\caption{\label{fig:nopump}Pumping parameters are set to zero value
to inspect the behavior of coherence and steady state
characteristic. The coherence $\rho_{bc}$, upper level population
$\rho_{aa}$, population inversion
$\rho_{aa}-\rho_{bb}$,$\rho_{aa}-\rho_{cc}$ are plotted in different
line style. Decay parameters are set to $\gamma_1=\gamma_2=1.0$
Initial conditions take as case$II$
:$\rho_{aa}=1,\rho_{bb}=0,\rho_{cc}=0,\rho_{bc}=\rho_{cb}=0 $,
pumping parameters are$r_1=r_2=0$ }.
\end{figure}

What will the system state like only by the action of decaying
process? It's suitable to take initial condition of case $II$ to
inspect this:
$\rho_{aa}=1,\rho_{bb}=0,\rho_{cc}=0,\rho_{bc}=\rho_{cb}=0 $, that
is, initially the upper state $|a\rangle$ is fully populated, while
the two lower states $|b\rangle$ and $|c\rangle$ are empty, the
coherence starts from zero value. If we cancel the pump field, the
system will then decays, without the disturbances of pumping field,
the atomic system will arrive at a positive maximal coherence steady
state, in which the upper level has no population, and the two lower
levels are equally populated with half of total
population:$\rho_{aa}=0,\rho_{bb}=\frac{1}{2},\rho_{cc}=\frac{1}{2}$;
and the coherence comes to the positive maximal value,
$\rho_{bc}=\frac{1}{2}$, as shown in Fig.\ref{fig:nopump}.

Interestingly, the final steady state is equal to the other initial
state in case $II$. The density matrix of this steady state take the
following form:
\begin{eqnarray}
\label{eq:weakmatrix} M=\left[
\begin{array}{cc}
 \rho_{bb} &\rho_{bc} \\
 \rho_{cb} &\rho_{cc}
   \end{array}
   \right]=\left[
   \begin{array}{cc}
 \frac{1}{2} &\frac{1}{2}\\
\frac{1}{2} & \frac{1}{2}
   \end{array}
   \right].
\end{eqnarray}
Similarly, if we takes the \emph{dual rail} representation, the
upper state can be expressed as:
$|\psi\rangle=\frac{1}{\sqrt{2}}(|01\rangle+|10\rangle)$. But this
state is not so stable as the "robust" state under the driving of
nonvanishing pumping field, as discussed in the upper text, we would
call it the "weak" steady state corresponding to the "robust" steady
state. This "weak" stability indicates the stochastic and incoherent
nature of spontaneous decay process, while the "robust" stability
shows the versatile applications of pumping field in level states
manipulation and overcoming the stochastic spontaneous. For example,
by adjusting the pumping parameters $r_1$ and $r_2$ according to the
decaying rates $\gamma_1$ and $\gamma_2$, we can generate steady
state from the "robust" steady to the "weak" steady state, which
will be very useful in quantum optics and quantum information
science.

\section{\label{sec:Diss} Discussion and Conclusion}
In the above, we have studied the level population transition and
the atomic coherence excitation process in the three-level $\Lambda$
atomic system in the joint action of incoherent pumping field and
vacuum reservoir, and find out that the system has rich coherence
and interference features like that in the action of coherent
pumping field, for example, the coherent population trapping (CPT)
state can be generated easily. So the upper system under discussion
will be very useful in applications to atomic coherence and
interference in quantum optics, such as lasing without inversion
(LWI), electromagnetically induced transparency (EIT) and quenching
spontaneous emission, and so on, which is very similar to that of
the $V$ type system\cite{PRA063829}. As a comparison, the CPT state
generated in $\Lambda$ system has better quality than that in the
$V$ type system, for it's unrelated to the decay parameters and only
relates to the pumping rates. With this better accessibility of the
$\Lambda$ system, it will deserve much more recognition in atomic
excitation study. But unfortunately, this kind of closely spaced
levels structure atomic system is difficult to get\cite{PRA063829},
and the experimental implementation of the upper attractive
applications seems to be a very difficult job. However the system we
are studying still shows interesting features in theoretical
research, for example, the two kind of extreme situations for
coherence value, i.e.,$\rho_{bc}=\pm\frac{1}{2}$ produce interesting
results: the "robust" steady state and the "weak" steady state,
which will be of great help to inspect the fine mechanism of atomic
coherence excitation, and to modify atomic states.

In conclusion, by solving the motion equations of density matrix
elements of a closed $\Lambda$ type three-level atomic system under
the interaction of incoherent pump field and decay process, we
derived a coherent population trapping steady state and multi-steady
states with different parameters sets for the atomic system. From
the solutions we showed a picture of the level population
transferring and revealed the mechanism of atomic coherence
excitation: the coherence is produced by the destructive
interference of incoherent pumping process and decay process. Owning
to different value of coherence, the system is able to generate
interesting states as the "robust" steady states and the "weak"
steady state, which promises fruitful potential applications in
quantum optics and in quantum information science.

\begin{acknowledgments}
 This work was supported by National Funds of Natural
Science (Grant No. 10504042).
\end{acknowledgments}

\bibliographystyle{unsrt}
\bibliography{reference} %
\end{document}